# Short-channel field effect transistors with 9-atom and 13-atom wide graphene nanoribbons

*Juan Pablo Llinas[1], Andrew Fairbrother[2], Gabriela Borin Barin[2], Wu Shi[4], Kyunghoon Lee[1], Shuang Wu[1], Byung Yong Choi[1,3], Rohit Braganza[1], Jordan Lear[1], Nicholas Kau[4], Wonwoo Choi[4], Chen Chen[4], Zahra Pedramrazi[4], Tim Dumslaff[6], Akimitsu Narita[6], Xinliang Feng[7], Klaus Müllen[6], Felix Fischer[5,8], Alex Zettl[4,8], Pascal Ruffieux[2], Eli Yablonovitch[1,8], Michael Crommie[4,8], Roman Fasel[2,9], Jeffrey Bokor[1]\**

[1]Dept. of Electrical Engineering and Computer Sciences, UC Berkeley, Berkeley, CA, USA

[2]Empa, Swiss Federal Laboratories for Materials Science and Technology, Dübendorf, CH

[3]Flash PA Team, Semiconductor Memory Business, Samsung Electronics Co. Ltd., Korea

[4]Dept. of Physics, UC Berkeley, Berkeley, CA, USA

[5]Dept. of Chemistry, UC Berkeley, Berkeley, CA, USA

[6]Max Planck Institute for Polymer Research, Mainz, DE

[7]Center for Advancing Electronics Dresden, TU Dresden, Dresden, DE

[8]Kavli Energy NanoSciences Institute at the University of California, Berkeley and the Lawrence Berkeley National Laboratory, Berkeley, CA, USA

[8]Dept. of Chemistry and Biochemistry, University of Bern, Freiestrasse 3, 3012 Bern, CH

*jbokor@eecs.berkeley.edu

**Bottom-up synthesized GNRs[1–8] and GNR heterostructures[9,10] have promising electronic properties for high performance field effect transistors (FETs)[11] and ultra-low power devices such as tunnelling FETs[12]. However, the short length and wide band gap of these GNRs have prevented the fabrication of devices with the desired performance and switching behaviour[13–16]. Here, by fabricating short channel ($L_{ch}$ ~20 nm) devices with a thin, high-κ gate dielectric and a 9-atom wide (0.95 nm) armchair GNR as the channel material, we demonstrate FETs with high on-current ($I_{on}$ >1 μA at $V_d$ = -1 V) and high $I_{on}/I_{off}$ ~$10^5$ at room temperature. We find that the performance of these devices is limited by tunnelling through the Schottky barrier (SB) at the contacts and we observe an increase in the transparency of the barrier by increasing the gate field near the contacts. Our results thus demonstrate successful fabrication of high performance short-channel FETs with bottom-up synthesized armchair GNRs.**

The electronic, optical and magnetic properties of GNRs can be engineered by varying their width and edge structure[17–19]. However, traditional methods to pattern GNRs, such as unzipping carbon nanotubes or lithographically defining GNRs from bulk graphene, yield GNRs with rough edges that degrade electronic transport[20]. Recent experiments have demonstrated bottom-up chemical synthesis of GNRs with uniform width and atomically-precise edges, in which the width and edge structure of the GNR is determined by the oligophenylene used in the polymerization step[1,2,5,9,10]. This synthetic uniformity produces GNRs with high structural and electronic homogeneity, which is required for integration of GNRFETs into large-scale digital circuits[21].

To create high performance GNRFETs, we used 9-atom and 13-atom wide armchair GNRs (9AGNRs and 13AGNRS, respectively). With a predicted band gap of 2.10 eV for the isolated 9AGNR and 2.35 eV for the 13AGNR[18], these are the narrowest band gap GNRs that have been synthesized on a surface with useful length for device fabrication (5AGNRs have a smaller band gap but are only ~10 nm long with current synthetic methods[7]). To synthesize the GNRs, the requisite monomer was evaporated onto a Au(111) surface under ultra-high vacuum and heated until it polymerized. Heating the substrate further causes individual polymers to planarize into GNRs (cyclodehydrogenation). The high quality of the GNRs is verified by high-resolution scanning tunnelling microscope (STM)[1,2,4] imaging as shown in Fig. 1a,b.

Fabrication of GNRFETs requires the transfer of GNRs from the Au growth surface onto an insulating surface and subsequent device fabrication steps, as shown in Fig. 1c and as described in the Methods. Unfortunately, standard imaging techniques (atomic force microscopy, scanning electron microscopy, transmission electron microscopy, etc.) were not useful in imaging single GNRs on insulating surfaces due to the GNR's small dimensions (~30 nm long, ~1 nm wide and <1 nm thick). Instead, we used Raman spectroscopy in order to verify that the structural integrity of the GNRs is maintained throughout the transfer and device fabrication process. As shown in Fig. 2a, the Raman spectrum with 785 nm wavelength excitation of the processed 9AGNRs looks identical to the spectrum taken of the as-grown 9AGNRs on Au. The presence of the radial breathing-like mode (RBLM) peak (311.5 cm$^{-1}$) is evidence that the GNR width and edge structure is intact throughout device processing[22,23]. Unlike 9AGNRs, the RBLM is not visible for the 13AGNR spectrum for either 532 nm or 785 nm excitation wavelengths due to off-resonance of the excitation (Fig. 2b and Fig. S2). Still, the 13AGNRFETs were processed with the same fabrication steps as the 9AGNRFETs and both types of devices exhibit similar transport characteristics (Fig. 3).

First, we fabricated devices with a nominal 20 nm channel length and a 50 nm $SiO_2$ gate dielectric as illustrated in Fig. 3a. Using the same fabrication methods we made two different types of samples: one with 9AGNRs and one with 13AGNRs. After patterning ~300 pairs of electrodes in the transferred GNR area, each defined channel was biased and tested for gate modulation of the current to find devices bridged by a GNR. Of the 300 devices, 28 devices and 29 devices were successfully fabricated for 9AGNR and 13AGNRs, respectively. This ~10% ratio of bridged contacts to open devices indicates that almost all of the devices found contain one GNR in the channel as demonstrated by Fig. S1.

These 9AGNRFETs and 13AGNRFETs, as shown in Fig. 3 and Fig. S3, showed similar electrical behaviour due to their similar band gap. The presence of a SB at the Pd-GNR interface is evident by the non-linear behaviour at low bias in the $I_d$–$V_d$ characteristics, shown in Fig. 3a,b. To determine the contributions of thermionic vs tunnelling current across the SB, we measured the devices in vacuum at 77 K, 140 K, 210 K and 300 K. As demonstrated in Fig. 3 c,d, there is no significant change in the characteristics at these different temperatures for either 13AGNRFETs or 9AGNRFETs and the off-state current is at the gate leakage level (Fig. S4). The weak temperature dependence in the current-voltage characteristics suggests that the limiting transport mechanism is tunnelling through the barrier as opposed to thermionic emission over the barrier at the contacts. Furthermore, the ambipolar behaviour observed at low temperatures is only realistically possible with tunnelling contacts, since thermally activated current is suppressed for electrons in a semiconductor with a band gap of >2 eV. Tunnelling contacts with weak temperature dependence have been observed for carbon nanotube FETs and other low-dimensional materials and verified via simulations[24–26]. Yet, the $I_{on}$ = ~100 nA in the devices shown in Fig. 3 is too low for high-performance applications, so the transmission through the SBs must be enhanced to improve the current.

Ionic liquid (IL) gating has been previously used to improve the transparency of the SBs in $MoS_2$[27]. Thus, we used the IL N,N-diethyl-N-(2-methoxyethyl)-N-methylammonium bis(trifluoromethylsulphonyl-imide (DEME-TFSI) to improve the electrostatic coupling between the gate electrode and the GNR channel, increase the field at the Pd-GNR interface and improve the transmission through the barriers. The $I_d$–$V_g$ behavior of a 9AGNRFET with IL gating is shown in Fig. 4b. This device shows clear enhancement in the on-current to ~200 nA at -0.2 V drain bias (as opposed to 3 nA at -0.4 V for the 50 nm $SiO_2$ dielectric device presented in Fig 4a). The transistor also switches at smaller gate voltages due to the high gate efficiency of the IL.

Since solid dielectric gates must be used for logic device applications, we fabricated scaled 9AGNR devices with a thin $HfO_2$ gate dielectric (effective oxide thickness of around 1.5 nm) as shown in Fig. 5. Resembling the IL device, the local $HfO_2$ back gate is more efficient at improving transmission through the SB than the thick $SiO_2$ global back gate[28]. As demonstrated by the $I_d$ vs $V_g$ shown in Fig. 5, the device exhibits excellent switching characteristics, $I_{on}/I_{off}$ ~$10^5$, and a high $I_{on}$ ~1 µA at $V_d$ = -1 V. This corresponds to a GNR-width (0.95 nm) normalized current drive of ~1000 µA/µm at -1 V drain bias, superior to previously reported top-down GNR devices[29–31]. Therefore, the scaled device structure with the improved gate efficiency allows for ultra-narrow bottom-up GNRs to outperform the narrow band gap top-down GNRs by mitigating the impact of the SBs on the contact resistance.

We thus successfully demonstrate high performance short-channel FETs with bottom-up synthesized armchair GNRs. These GNRFETs have excellent switching behaviour and on-state performance after aggressively scaling the gate dielectric. Bottom-up GNR devices are therefore good candidates for high-performance logic applications, especially with advances in densely aligned GNR synthesis[32] as well as narrow band gap GNR growth[7]. Our methodology can be applied to other exotic device structures as well, such as tunnel FETs, which incorporate atomically precise GNR heterostructures[9,10,12].

METHODS

**9AGNR growth.** 9AGNRs are synthesized from 3',6'-dibromo-1,1':2',1"-terphenyl precursor monomers.[4] First, the Au(111)/mica substrate (200 nm Au; PHASIS, Geneva, Switzerland) is cleaned in ultra-high vacuum by two sputtering/annealing cycles : 1 kV $Ar^+$ for ten minutes followed by a 470 °C anneal for ten minutes. Next, the monomer is sublimed onto the Au(111) surface at a temperature of 60-70 °C, with the substrate held at 180 °C. After 2 minutes of deposition (resulting in approximately half monolayer coverage), the substrate temperature is increased to 200 °C for ten minutes to induce polymerization, followed by annealing at 410 °C for ten minutes in order to cyclodehydrogenate the polymers and form 9-AGNRs.

**13AGNR growth.** 13AGNRs were synthesized using 2,2'-Di((1,1'-biphenyls)-2-yl)-10,10'-dibromo-9,9'-bianthracene building blocks.[2] Similar to the 9AGNR substrate, the Au(111)/mica substrate (200 nm Au; PHASIS, Geneva, Switzerland) is cleaned in ultra-high vacuum by two sputtering/annealing cycles : 1 kV $Ar^+$ for ten minutes followed by a 450 °C anneal for ten minutes. The monomer was sublimed at 222 °C onto the clean substrate held at room temperature. The sample was then slowly annealed stepwise to 340 °C to form 13AGNRs.

**Preparation of 50 nm $SiO_2$ back gates.** Using dry oxidation, 50 nm $SiO_2$ was grown on heavily doped 150 mm silicon wafers. Alignment markers and large pads for electrical probing were patterned using standard photolithography and lift-off patterning of 3 nm Cr and 25 nm Pt. The wafer was then diced and individual chips were used for GNR transfer and further device processing.

**Preparation of 6.5 nm $HfO_2$ local back gates.** Using dry oxidation, 100 nm $SiO_2$ was grown on 150 mm silicon wafers. The local back gates were lithographically patterned and dry etched into the SiO$_2$ followed by lift-off of 3nm Ti and 17 nm Pt.[33] 6.5 nm HfO$_2$ was grown in an atomic layer deposition system at 135 °C. Alignment markers and large pads for electrical probing were patterned using standard photolithography and lift-off of 3 nm Cr and 25 nm Pt. The wafer was then diced and individual chips were used for GNR transfer and further device processing.

**GNR transfer and patterning of source-drain electrodes.** GNR/Au/mica was floated in 38% HCl in water, which caused the mica to delaminate with the Au film remaining floating on the surface of the acid.[10] The floating gold film was picked up with the target substrate, with the GNRs facing the dielectric surface. Subsequent gold etching in KI/I$_2$ yielded isolated, randomly distributed GNRs with sub-monolayer coverage on the target substrate. After the GNR transfer, poly-methyl methacrylate (PMMA, molecular weight 950K) was spun on the chips at 4 Krpm and followed by a 10 min bake at 180 °C. ~300 source drain electrodes (100 nm wide, 20 nm gaps) were patterned using a JEOL 6300-FS 100 kV e-beam lithography system and subsequently developed in 3:1 IPA:MIBK at 5 °C. 10 nm Pd was deposited using e-beam evaporation and lift-off was completed in Remover PG at 80 °C.

**Raman characterization**. Raman characterization of the 9AGNR was performed with a Bruker SENTERRA Raman microscope using a 785 nm diode laser with 10 mW power and a 50x objective lens, resulting in a 1-2 micrometer spot size. No thermal effects were observed under these measurement conditions and an average of 3 spectra from different points was made for each sample. Raman measurements of the 13AGNR were made with a Horiba Jobin Yvon LabRAM ARAMIS Raman microscope using 532 nm and 785 nm diode lasers with 10 mW power each and a 50x objective lens, resulting in a 1-2 micrometer spot size. No thermal effects were observed under these measurement conditions and an average of 5 spectra from different points was made for each sample.

**Electrical characterization.** Devices were first screened in ambient conditions using a cascade probe station and an Agilent B1500A parameter analyser. Vacuum and variable temperature measurements were then performed in a Lakeshore probe station. Ionic liquid devices were measured with a $V_g$ sweep speed of 50 mV/s.

ACKNOWLEDGMENT

This work was supported in part by the Office of Naval Research BRC program under Grant N00014-16-1-2229, DARPA, the U. S. Army Research Laboratory and the U. S. Army Research Office under contract/grant number W911NF-15-1-0237, the Swiss National Science Foundation, the DFG Priority Program SPP 1459 and Graphene Flagship (No. CNECT-ICT-604391). Work at the Molecular Foundry was supported by the Office of Science, Office of Basic Energy Sciences, of the U.S. Department of Energy under Contract No. DE-AC02-05CH11231. Additional support was received from the Director, Office of Science, Office of Basic Energy Sciences, Materials Sciences and Engineering Division, of the U.S. Department of Energy under Contract No. DE-AC02-05-CH11231, within the sp2-Bonded Materials Program (KC2207), which provided for development of the IL gating method. J.P.L. is supported by the Berkeley Fellowship for Graduate Studies and by the NSF Graduate Fellowship Program.

AUTHOR CONTRIBUTIONS

J.P.L., B.Y.C., R.B. and J.B. fabricated and measured the devices on $SiO_2$. J.P.L., J.L., K.L., S.W., J.B. and E.Y. fabricated and measured the devices on $HfO_2$. W.S. and A.Z. performed the IL gating experiments. A.F, G.B.B, P.R. and R.F. performed growth, transfer, STM, and Raman spectroscopy of 9AGNRs. N.K., W.C., C.C., Z.P. and M.C. performed growth and STM measurements of 13AGNRs. J.P.L. and J.B. transferred and performed Raman spectroscopy of 13AGNRs. F.F. synthesized the 13AGNR precursor molecule. T.D., A.N., X.F. and K.M. synthesized the 9AGNR precursor molecule. All the authors discussed and wrote the paper.

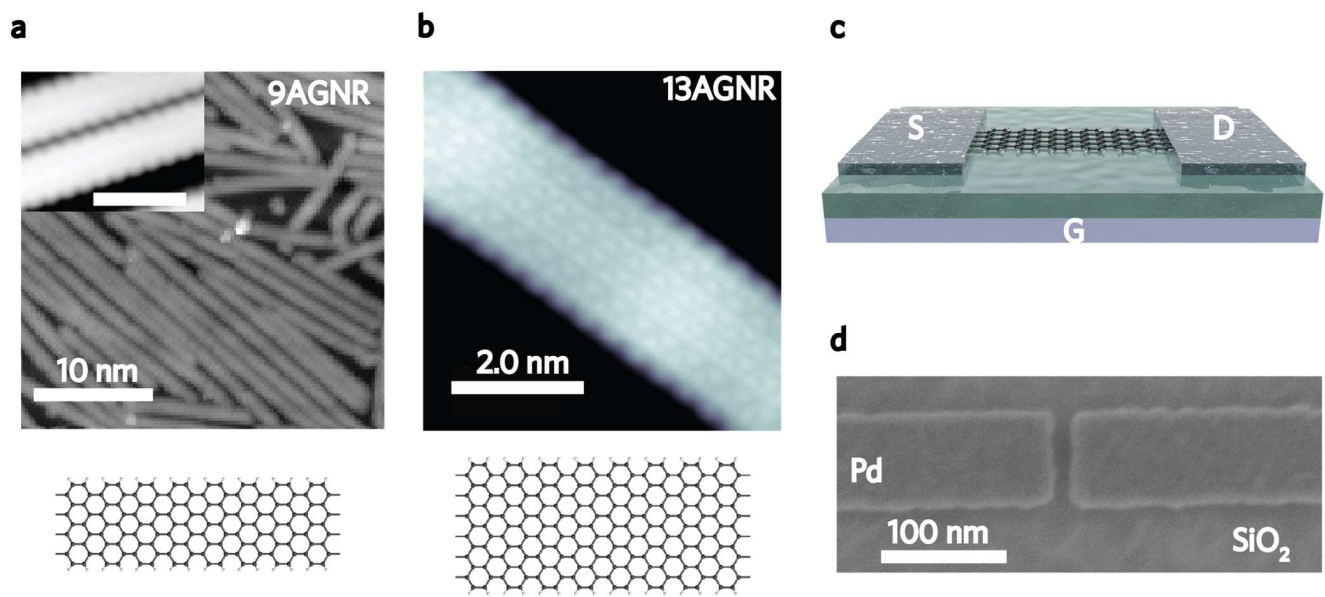

**Figure 1 High resolution STM GNR characterization and FET structure** (a) STM image of synthesized 9AGNR on Au ($V_s$ = 1 V, $I_t$ = 0.3 nA). Inset: High resolution STM image of 9AGNR on Au ($V_s$ = 1 V, $I_t$ = 0.5 nA) with a scale bar of 1 nm (b) High resolution STM image of 13AGNR on Au ($V_s$ = -0.7 V, $I_t$ = 7 nA). (c) Schematic of the short channel GNRFET with a 9AGNR channel and Pd source-drain electrodes (d) Scanning electron micrograph of the fabricated Pd source-drain electrodes.

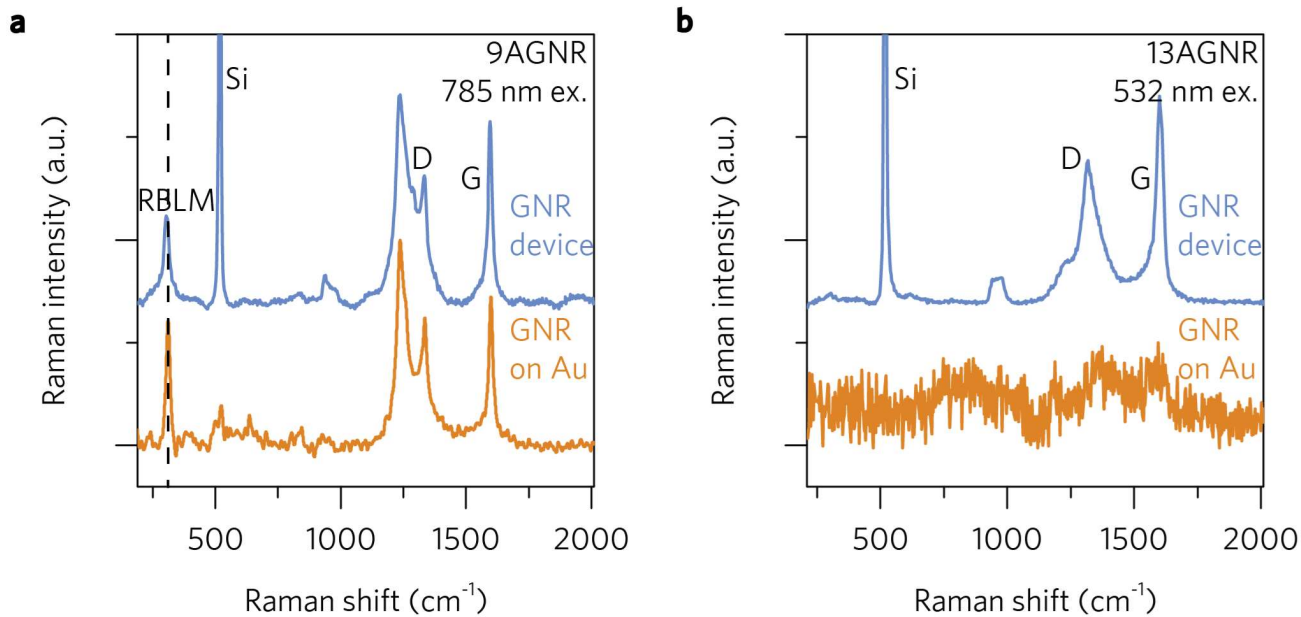

**Figure 2 Raman spectra of as-grown GNRs on Au and GNRs after transfer and device processing.** Raman spectra of (a) 9AGNRs and (b) 13AGNRs on the Au(111) growth substrate and after device fabrication shows that the GNRs remain intact. Since the excitation is off-resonance with the 13AGNR absorption, the Raman signal is weak on Au and the RBLM is not visible.

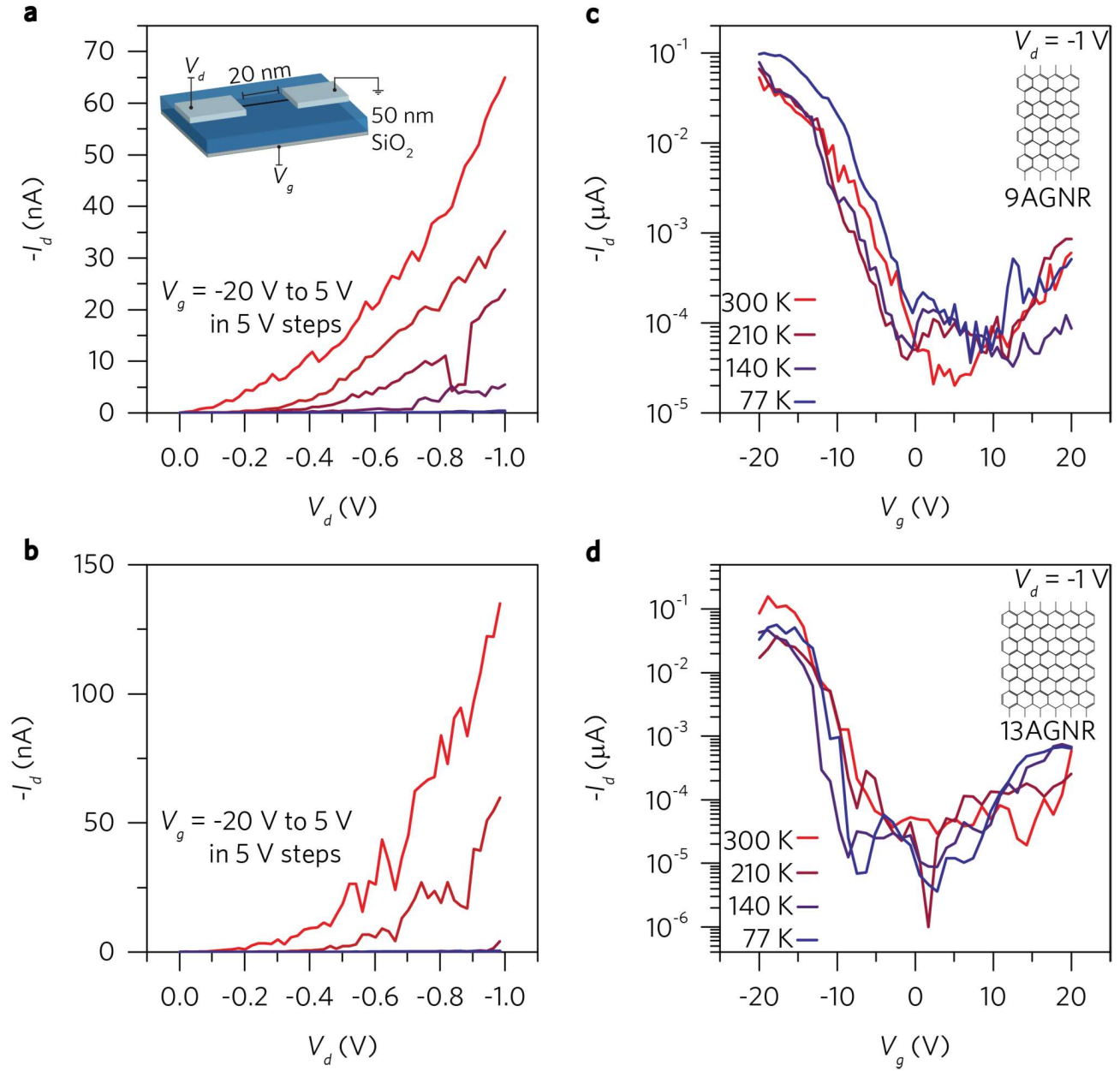

**Figure 3 Transport characteristics of 9AGNRs and 13AGNRs gated with 50 nm SiO$_2$ gate oxide.** The presence of a SB is confirmed by non-linear current behaviour at low drain bias and lack of current saturation at high drain bias for both (a) 9AGNRs and (b) 13AGNRs. The weak temperature dependence in the $I_d$-$V_g$ behaviour in (c) 9AGNRs and (d) 13AGNRs indicates that tunnelling through the Pd-GNR SBs is the limiting transport mechanism of the device.

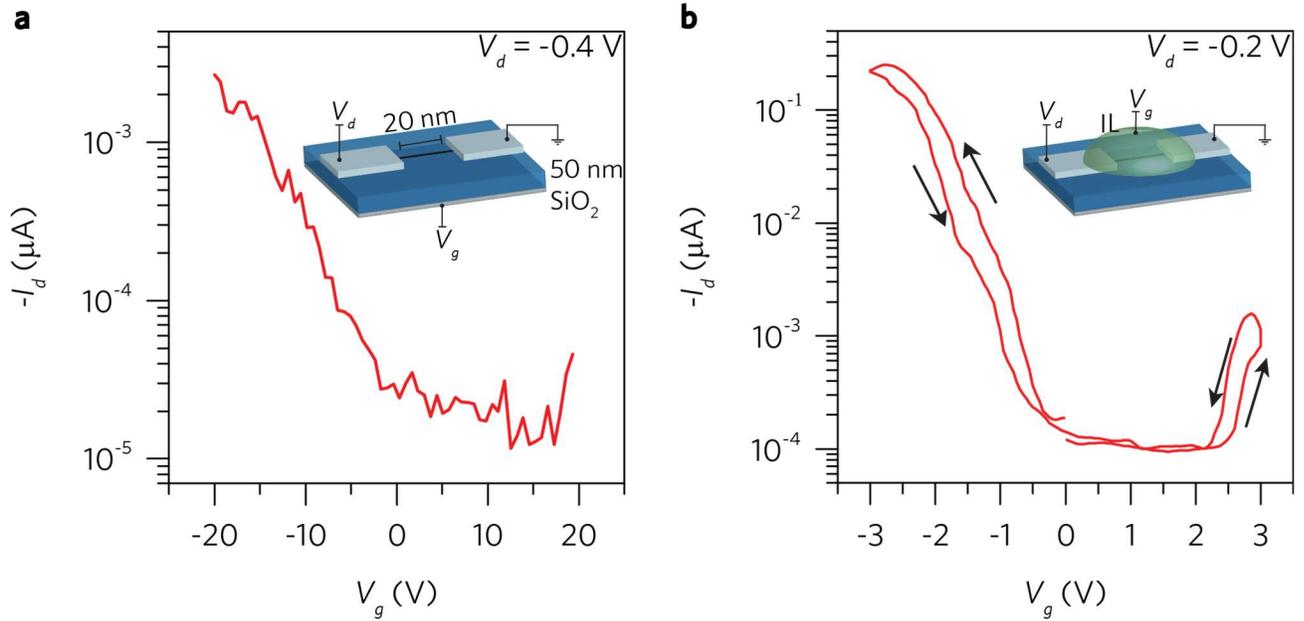

**Figure 4 Ionic liquid gating of a 9AGNRFET at room temperature.** (a) $I_d$-$V_g$ characteristics of the device gated by the thick 50 nm $SiO_2$ gate oxide (b) $I_d$-$V_g$ characteristics of the device gated with the ionic liquid which shows clear ambipolar behaviour and improved on-state performance. Inset: ionic liquid (DEME-TFSI) gated 9AGNRFET device schematic.

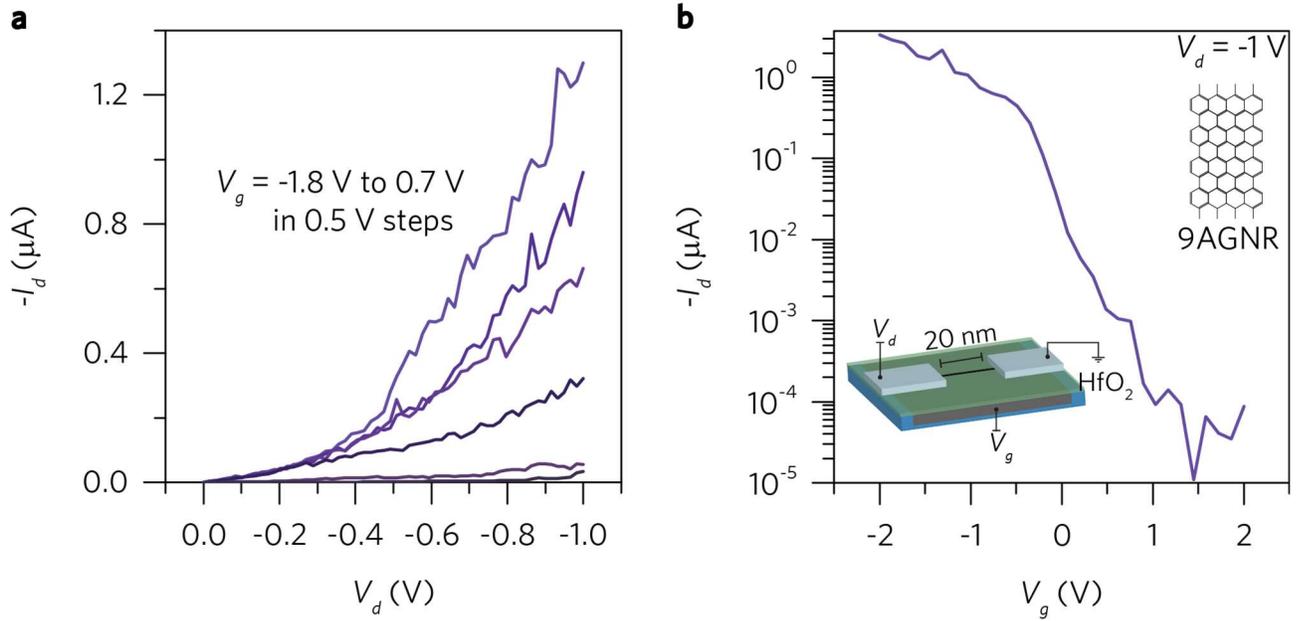

**Figure 5 Transport characteristics of a scaled, high performance 9AGNRFET at room temperature.** (a) $I_d$-$V_d$ characteristics of the scaled 9AGNRFET. (b) $I_d$-$V_g$ of the devices show high $I_{on}$ >1 µA for a 0.95 nm wide 9AGNR and high $I_{on}/I_{off}$ ~$10^5$. Inset: scaled 9AGNRFET schematic.

## Supplemental Information

**Extraction of number of GNRs in the channel**

We used a Monte Carlo simulation to estimate the number of GNRs in our device channels based on our device yield. Assuming a uniform spatial distribution of GNRs, we simulate the expected device yield and distribution of number of GNRs in the channel. The input parameters of the simulation were the GNR number density on the surface and GNR length. We varied these parameters to generate Fig. S1. With the experimentally obtained yield of ~10%, the percentage of devices with more than 1 GNR in the channel goes as high as 8% for higher surface density and 4% for low surface density. Out of the devices with multiple GNRs, an insignificant percentage has more than 2 GNRs/channel. Thus, we estimate that only 1-3 devices out of ~30 fabricated devices have 2 GNRs in the channel. However, it is unclear whether these devices would account of the high tail end of the on-current distribution since both GNRs would have to have good contact length under the Pd contacts to improve conduction over a single GNR channel with a large contact length.

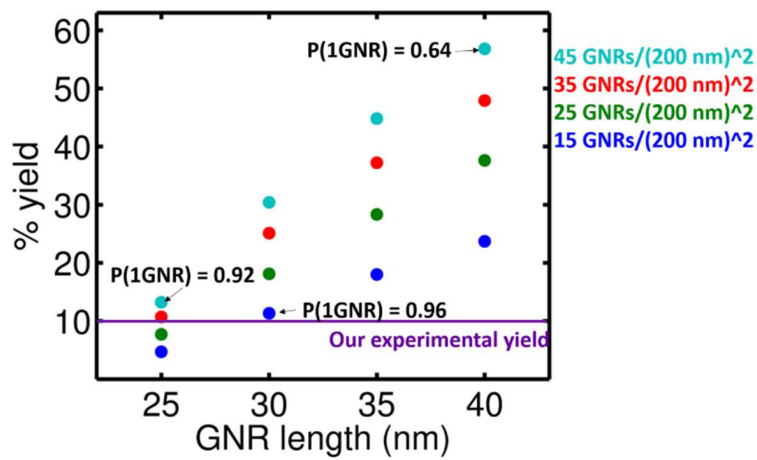

**Figure S3.** Simulated % yield of working devices as a function of GNR length and number density. The P(1GNR) values denote the probability of a yielded device to have a single GNR in the channel. With our ~10% experimental yield, we estimate that only 1-3 devices out of 30 contain 2 GNRs in the channel.

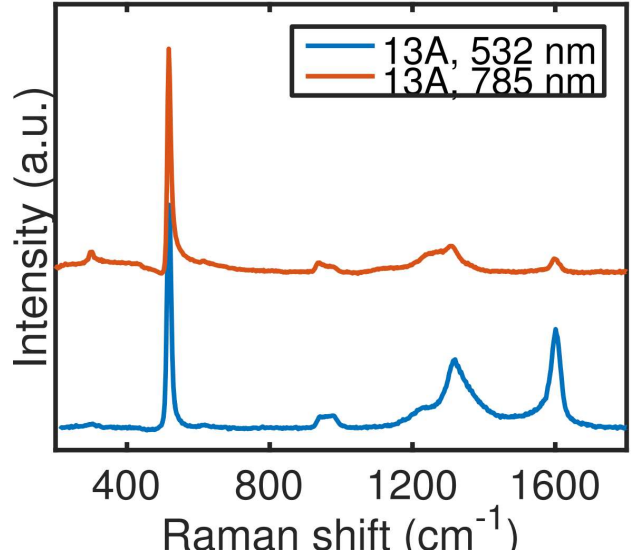

**Figure S2.** Raman spectra of 13AGNRs after device fabrication using 532 nm and 785 nm wavelength excitation. The RBLM is not detectable under these excitation conditions due to excitation off-resonance effects.

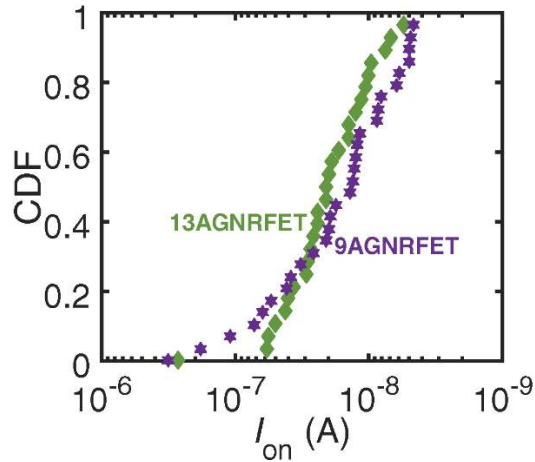

**Figure S3.** Cumulative distribution function (CDF) of $I_{on}$ in 13AGNRFETs and 9AGNRFETs with 50 nm $SiO_2$ gate dielectrics. The CDF is defined as the total fraction of devices with on-current greater than the given value of $I_{on}$. Both types of devices have similar behavior due to the similar band gap and variations in on-state performance are most likely due to variations in the overlap length between the Pd and GNR and variations in the channel length.

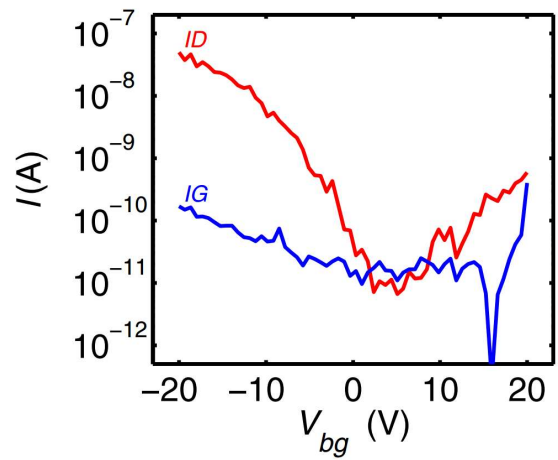

**Figure S4.** $I_d$-$V_{bg}$ characteristics of the 9AGNRFET shown in Fig 3 in the main text which shows that the gate leakage limits the off-current.